\documentstyle[12pt]{article}

\begin{document}

\begin{center}
{\Large\bf  Triton Binding Energy and 
Minimal Relativity} 
\\
\vspace*{0.5cm}
F. Sammarruca and R. Machleidt\\
 {\it Department of Physics,
  University of Idaho,\\
   Moscow, Idaho 83843, U.S.A.}
\\
\vspace*{.1cm}
\today
\end{center}
\vspace{1cm}

\begin{abstract}

For relativistic three-body calculations, essentially two different 
approaches are in use: field theory and relativistic direct interactions.
Results for relativistic corrections of the triton binding energy 
obtained from the two approaches differ even in their sign, which is 
rather puzzling. In this paper, we discuss the origin of such discrepancy. 
We show that the use of an invariant two-body amplitude, as done in the 
field-theoretic approach, increases the triton binding energy by 
about 0.30 MeV. This may explain a large part of the discrepancy. 

\end{abstract}


\pagebreak


 
In recent years, many high-precision, charge-dependent 
calculations of the triton binding energy 
with realistic nucleon-nucleon (NN) interactions have been performed. 
Employing {\it local} two-nucleon potentials of high quality, 
in a 34-channel charge-dependent non-relativistic Faddeev calculation, 
the triton binding energy is predicted to be 
7.62(1) MeV \cite{Friar93}.   
With a recent, high-precision version of the ({\it non-local}) Bonn potential
(CD-Bonn), the same type of Faddeev calculation predicts 8.00 MeV for the 
triton binding energy \cite{MSS96}, the difference in the predictions       
originating 
from variations in the off-shell behaviour of the potentials.

However, 
due to the presence of high momentum components in the triton wave function,
it is important to go beyond 
the non-relativistic approximation.                    
Moreover, even if relativistic effects 
turn out to be small, they may still be large 
as compared to the discrepancy with the experimental value, which is 8.48 MeV. 
Therefore, the present situation 
calls for an accurate knowledge of the effect of relativity 
on the three-nucleon bound-state.

Let us briefly 
 review  
 the current status of the relativistic three-body
bound-state problem.
Essentially, one may distinguish between two major theoretical frameworks:
\begin{enumerate} 
\item field theory, 
\item relativistic direct interactions.
\end{enumerate}

Within the first scheme, Rupp and Tjon \cite{RT92} 
have used Bethe-Salpeter (BS) equations
for two and three particles. 
They constructed covariant multirank separable interactions
from phenomenological as well as meson-exchange potentials (Paris and Bonn).
They predict 0.29-0.38 MeV more binding with respect to  non-relativistic 
results.

The field-theoretic approach has also been pursued in our  own work, in 
which we apply 
the relativistic three-dimensional version of the Bethe-Salpeter
equation proposed by Blankenbecler and Sugar (BbS), together with relativistic
one-boson exchange potentials \cite{SXM92}. 
We find an increase of the triton binding energy due to relativistic effects
of 0.19 MeV using the Bonn~B potential \cite{SXM92}, and obtain the same result
with the more recent CD-Bonn
\cite{MSS96}. 

 Gross and collaborators have recently reported 
results for the triton binding energy very close to the experimental 
value \cite{Gross96}, using a relativistic field-theoretic framework.

Alternative to field theory, other approaches have been pursued, in 
which the attempt is made 
to combine quantum mechanics with the requirements of relativistic 
invariance. Within the scheme of relativistic direct interactions, 
relativistic Hamiltonians are defined as the 
sum of relativistic one-body kinetic energies, two- and many-body interactions
and their boost corrections. The latter is derived from commutation relations
of the Poincar\'e group \cite{Bak52, Foldy61, Keist91, Keist95}.
Recent work within this approach \cite{Carls93} reports 
that relativistic effects 
{\it reduce} the triton binding energy by 0.34 MeV.

In summary, 
 even the sign of the relativistic correction is controversial. This 
is  rather worrisome, and 
  calls for further 
investigation of the many facets of the problem.   
It is the purpose of this note to discuss some aspects involved.

Within the framework of relativistic quantum field theory, there are 
essentially three sources of relativistic effects in the three-body system:
\begin{itemize}
\item The use of an invariant two-nucleon amplitude 
as input to the three-body calculation, 
\item relativistic kinematics, 
\item relativistic Faddeev equations/propagators. 
\end{itemize} 

For best transparency of the investigation, 
it helps to single out each of these effects, with the first one being the 
focus of this note.  

To investigate this point, we use a simple prescription, known as 
{\it minimal relativity} \cite{BJK69}.                             
 A relativistic (invariant) two-body
$t$-matrix, $t_{rel}$, can be related to the non-relativistic $t$-matrix, 
$t_{NR}$ (solution of the Lippman-Schwinger equation), by 
\begin{equation}
t_{rel}({\bf q'},{\bf q}) = \sqrt{\frac{E'}{m}} t_{NR}({\bf q'},{\bf q}) 
\sqrt{\frac{E}{m}}
\end{equation}
with $E=\sqrt{m^{2} + q^{2}}$ and $E' = \sqrt{m^{2} + q'^{2}}$
($q\equiv |{\bf q}|$,
$q'\equiv |{\bf q'}|$).
When $t_{rel}$
of Eq.~(1) is inserted into the relativistic BbS equation 
(together with an 
analogous expression relating $V_{rel}$ and $V_{NR}$), 
the usual non-relativistic Lippmann-Schwinger equation 
for $t_{NR}$ is obtained.

Applying a variety 
 of NN potentials, we find that the use of 
$t_{rel}$ as defined in Eq.~(1) increases 
the triton binding energy by 0.25-0.37 MeV, see Table 1.
Note that Eq.~(1) allows to calculate the effect of 
minimal relativity for any potential (even non-relativistic ones)  
with no need to refit the nucleon-nucleon phase parameters~\cite{foot1}.

Of course, the other aspects of relativity mentioned above also 
 play a role, but they introduce uncertainties of a different nature. For
instance, various three-dimensional 
relativistic three-body equations are available, which all have in 
common the elimination of the relative time component in the original 
four-dimensional integration over internal momenta. This procedure is carried
out in such a way as to preserve relativistic invariance and three-particle
unitarity, but it is not unique. Also, the effect of relativistic kinematics
is dependent on the way 
relative momenta of the interacting pair are defined \cite{SG95}. 
Thus, 
in the 
field-theoretic approach, some ambiguities enter when a relativistic
three-dimensional reduction of the BS equation is used. These 
require a separate investigation. 
On the other hand, 
 the effect of using an invariant two-body amplitude in the
Faddeev equations, which we have chosen to single out in this note,
can be estimated independently by means of the minimal relativity
prescription, Eq.~(1).
This effect is absent
from approaches based on relativistic direct interactions,
since they are not manifestly covariant. 
Therefore, it may shed some light on the discrepancy between the
results from the two 
major theoretical frameworks. 

The present situation can then be summarized as follows: 
\begin{itemize}
\item 
Within a non-relativistic framework, and with local potentials, 
the triton binding energy is predicted to be 7.6 MeV, with no disagreement 
among different calculations. 
\item 
 On the other hand, when relativity is included, 
one group \cite{Carls93} reports
less binding, while others find  more \cite{RT92, SXM92, Gross96}. 
The latter                 
 approaches are all based on relativistic quantum field-theory
and obtain more binding energy regardless the dynamical input.              
Minimal relativity may explain the major reason for the extra binding. 

\end{itemize}

Thus, a rather puzzling picture emerges: relativity increases drastically
the uncertainty in the predictions, which now ranges from 7.3 to 8.2 MeV.

This state of affairs 
 is unacceptable, and opens a much more complex 
and fundamental issue, 
namely, how to define a relativistic correction \cite{Keist91}.
There is actually no unique way 
to define a {\it relativistic correction}, because
there is no unique way to define a non-relativistic limit. Equivalent 
covariant theories may differ in their non-relativistic limit, since the 
latter depends on how such limit is taken \cite{Keist91}.
So, strictly speaking, the only safe way to proceed is to start from 
a covariant theory.

The field-theoretic approaches we have mentioned above are consistently
relativistic from the outset, while the calculations using
direct interactions~\cite{Carls93} are
based on a non-relativistic baseline, with relativistic 
corrections applied in the 
form of a $p/m$ expansion. 
As investigated by Gl\"ockle and coworkers~\cite{GM81},
such an expansion may be risky and hard to control 
and, therefore, could be the source for 
the observed repulsion.

This work was supported in part by the U.S.\ National
Science Foundation under Grant No.\ PHY-9211607.

\pagebreak

\begin{table}
\caption{{\bf Effect of minimal relativity on the triton binding energy 
calculated for a variety of recent high-precision NN potential models.}}
\begin{tabular}{cc}
   \\
\hline\hline
NN Potential Model &\hspace*{1.5cm} Increase of Triton Binding Energy (MeV)\\
\hline
CD-Bonn [2]   &  0.25\\
Nijmegen-I [12] & 0.27\\
Nijmegen-II [12] & 0.37\\
Reid'93 [12] & 0.37\\
\hline\hline
\end{tabular}
\end{table} 

\begin{thebibliography}{99}
\bibitem{foot1} For an uncoupled partial wave,
the phase shift $\delta$  is related to the invariant scattering
amplitude, $t_{rel}$, by
\begin{displaymath}
e^{i\delta}\sin\delta = -\frac{\pi}{2}\, q\, \frac{m^2}{E}\, t_{rel} \; ,
\end{displaymath}
while for the nonrelativistic $t$-matrix, $t_{NR}$,
the relationship reads:
\begin{displaymath}
e^{i\delta}\sin\delta = -\frac{\pi}{2}\, q\, m\, t_{NR} \; .
\end{displaymath}
\end{thebibliography}
\end{document}